\documentstyle[12pt]{article}
\input{psfig}
\begin{document}
\newcommand {\ber} {\begin{eqnarray*}}
\newcommand {\eer} {\end{eqnarray*}}
\newcommand {\bea} {\begin{eqnarray}}
\newcommand {\eea} {\end{eqnarray}}
\newcommand {\beq} {\begin{equation}}
\newcommand {\eeq} {\end{equation}}
\newcommand {\state} [1] {\mid \! \! {#1} \rangleg}
\newcommand {\sym} {$SY\! M_2\ $}
\newcommand {\eqref} [1] {(\ref {#1})}
\newcommand{\preprint}[1]{\begin{table}[t] 
           \begin{flushright}               
           \begin{large}{#1}\end{large}     
           \end{flushright}                 
           \end{table}}                     
\def\Acknowledgements{\bigskip  \bigskip {\begin{center} \begin{large}
             \bf ACKNOWLEDGMENTS \end{large}\end{center}}}

\newcommand{\half} {{1\over {\sqrt2}}}
\newcommand{\dx} {\partial _1}

\def\Dslash{\not{\hbox{\kern-4pt $D$}}}
\def\cmp#1{{\it Comm. Math. Phys.} {\bf #1}}\begin{titlepage}
\titlepage
\rightline{TAUP-2412-97}
\vskip 1cm
\centerline{{\Large \bf Generalized Casimir Energies for Systems }}
\centerline{{\Large \bf with Arbitrary ($\epsilon(z)$) Planar Geometry}}
\vskip 1cm

\centerline{O. Kenneth}

\vskip 1cm
\begin{center}
\em School of Physics and Astronomy
\\Beverly and Raymond Sackler Faculty of Exact Sciences
\\Tel Aviv University, Ramat Aviv, 69978, Israel
\end{center}
\vskip 1cm
\begin{abstract}
The Casimir force between parallel plates of arbitrary kind is shown to be simply related to the plates transmission and reflection coefficient. A trivial application of this general relation leads to known Lifshitz force between dielectrics as well as its generalizations.  
\end{abstract}
\end{titlepage}

The simple two ideally conducting parallel plates Casimir force $F=-{\pi^2\over 240}{\hbar c\over a^4}$\cite{cas1} recently been measured experimentally\cite{lam,mjh}. It has various generalizations to different kinds of fields, boundary conditions and geometries whose calculation is in general quite complicated\cite{boy,sch,cyl}\cite{rr}. A new variant in which each of the plates conducts in specific in general different direction has been exactly evaluated recently\cite{11} and can hopefully be seen experimentally\cite{n2}. Another generalization is to replace the ideal conducting plates by dielectrics. Thus we may for example consider the vacuum energy corresponding to electromagnetic action in matter ${1\over 8\pi}\int d^4x (\varepsilon E^2-{1\over\mu}B^2)$ as some very complicated functional of $\varepsilon(x),\mu(x)$. In general this energy will contain divergent terms. When we consider the energy difference between two configurations (given by $\varepsilon(x),\mu(x)$) the divergent terms should cancel provided that the configurations may be related by a physical process which does not involve a change in internal matter properties (otherwise such a change would bring about a dependence on the physical UV cutoff). In particular relative displacement of the different dielectrics in space should result in a (finite) force.
The special case of the force between two infinite parallel dielectrics (corresponding to $\varepsilon={\left\{\begin{array}{cc} \epsilon_1&  z<0 \\ 1& 0<z<a\\ \epsilon_2 &a<z \end{array}\right.}$) was considered long ago by Lifshitz\cite{lif}.

We shall consider in the following general configurations depending only on the z coordinate. We shall show using path integral techniques that in this case the vacuum Casimir energy may be simply related to the transmission coefficient through the system This reduces the computation of the force to a solution of an elementary one dimensional scattering problem. In particular to find the electromagnetic Casimir force one merely need to know the classical reflection and transmission coefficient through the relevant system. In fact the relation between the Casimir energy and the transmission coefficient follows from of the relation between scattering amplitudes singularities and the eigenenergies.

We use the functional integral formalism to write the Casimir energy as $e^{-ET}=\int{\cal D}{\phi}e^{-S}=\det(...)^{-{1\over 2}}$.
Consider first a scalar field described by an action $S=\int d^4x (\phi\Box\phi+\varepsilon(z)\phi^2)$. To compute the functional determinant of this we first Wick rotate and Fourier transform in the three directions in which there is a translation symmetry-$t,x,y$.
After doing that we are left with the action $\int d^3k\int dz(-\phi\partial_z^2\phi+(\varepsilon(z)+k^2)\phi^2)$. The total Casimir energy for this problem is given by summing contributions corresponding to all possible values of $\vec{k}\equiv(k_t,k_x,k_y)$.

 $E={1\over 2}A\int{d^3k\over(2\pi)^3}\log\det{\cal H}_k$ where $\det{\cal H}_k$ is the functional determinant of the one dimensional schr\"{o}dinger like operator ${\cal H}_k=-\partial^2+\varepsilon+k^2$. 

Discretizing this using $\int dz \phi(k^2-\partial_z^2+V(z))\phi\sim\sum\left( (k^2+V(z_i))\phi_i^2-\phi_i{\phi_{i+1}+\phi_{i-1}-2\phi_i\over\Delta z^2}\right)\Delta z$
leads us to the determinant
$$\hspace{-1cm}I_n={\left |\begin{array}{ccccccccc} (k^2+V(z_1))\Delta z^2+2& -1&0&.&.&.& & & \\ -1& (k^2+V(z_2))\Delta z^2+2& -1 & & & & & &    \\ 0& -1&\hspace{1cm} & & & & & &  \\ .& & & & & & & &. \\ .& & & & & & & &. \\ .& & & & & & & &. \\  .& & & & &.&.&.&(k^2+V(z_n))\Delta z^2+2\end{array}\right |}$$
The determinant easily $I_n$ is seen to satisfy a recursion relation $I_n=((k^2+V(z_n))\Delta z^2+2I_{n-1}-I_{n-2}$\newline which  may also be written in the form ${I_n+I_{n-2}-2I_{n-1}\over\Delta z^2}=(k^2+V(z_n))I_{n-1}$\newline hence in the continuum limit  $I''-(k^2+V)I=0$\newline
i.e. the determinant satisfies the corresponding schr\"{o}dinger equation.

This is in complete accordance with another well known trick\cite{col} for calculating determinants of one dimensional schr\"{o}dinger operators.

 We are assuming that $\varepsilon(z)$ vanishes at $z=\pm\infty$. 
Under this assumption we know that the solutions of ${\cal H}\psi=0$ behave asymptotically as $\psi\sim e^{\pm kz}$. Let us look for a solution which is regular at minus infinity i.e. $\psi\rightarrow e^{kz}$ for $z\rightarrow -\infty$. Such a solution always exist and is unique .However in general it will not be regular at $+\infty$. Rather it behaves in this limit as $\psi\rightarrow Ae^{kz}+Be^{-kz}$. The function $\psi$ is a regular solution of ${\cal H}\psi=0$ if and only if the constant $A(k)=limit_{z\rightarrow\infty}(\psi(z)e^{-kz})$ is zero. On the other hand we know that a regular solution (which is necessarily given by the above defined $\psi$) exists if and only if $\det{\cal H}=0$. This suggests the identification  $\det{\cal H}=\lim(\psi(z)e^{kz})$. 

A more rigorous treatment starts from a similar problem where the particle is confined to a finite segment $[z_1,z_2]$ of length $L$ with boundary conditions $\psi(z_1)=\psi(z_2)=0$. In this case we may define a function $\psi(z)$ to be the solution of  ${\cal H}\psi=0$ satisfying the boundary conditions $\psi(z_1)=0,{\psi}'(z_1)=1$ and we easily see that $\psi(z_2)=0\Leftrightarrow\det{\cal H}=0$. 
The advantage of dealing first with the problem on a finite segment is that now the argument is correct for complex valued k while the previous argument was justified only if ${\it Re}(k)>0$ (while we know that most eigenvalues of ${\cal H}$ have in fact  ${\it Re}(k)=0$). Hence we may now conclude that $\psi_k(z_2)$ and $\det{\cal H}_k$ have exactly the same zeroes all over the complex plane. We also know from general properties of schr\"{o}dinger operator in one dimension that all the zeroes of  $\det{\cal H}$ are simple since all eigenvalues are nondegenerated.As $k\rightarrow\infty$ the problem becomes classical and one should be able to deduce the behaviour of both  $\psi_k(z_2)$ and of $\det{\cal H}_k$ at that limit from semi classical considerations. This allows concluding that $\psi_k(z_2)$ and  $\det{\cal H}_k$ are equal up to an ($\varepsilon(x)$-independent) normalization constant.

We are interested in knowing $\det{\cal H}$ only for real positive k. For such k the boundary conditions  $\psi(z_1)=0$ may be safely replaced by demanding $\psi\rightarrow e^{kz}$ for $z\rightarrow -\infty$. Making this change just amounts to neglecting a term of the order $e^{-kL}$ which disappears once when $L\rightarrow\infty$. By the same reason we may also replace  $\psi(z_2)$ by the coefficient of the increasing exponential $e^{kz}$ in the asymptotic expression for $\psi(z)$ (at this step we also renormalized the determinant by the constant factor $e^{kL}$ which is legitimate). It is also easy to see that had we used different boundary conditions such as $\psi'(z_1)=0$ instead of  $\psi(z_1)=0$ the result would only change by something proportional to $e^{-kL}$.

Given any function $\varepsilon(z)$ we may according to the above compute $\det{\cal H}_k$ by solving (possibly numerically) a one dimensional schr\"{o}dinger equation and then find the Casimir energy from the relation $E={1\over 2}A\int{d^3k\over(2\pi)^3}\log\det{\cal H}_k$.
We saw that $\det{\cal H}_k$ is given by the coefficient of $e^{kz}$ in the expression of $\psi$ at the right of the potential(i.e. for $z\gg 1$) when we assume that left of the potential (i.e. $z\ll -1$) $\psi$ was given by purely  $e^{kz}$. Analytically continuing this to imaginary values of k writing $k=-iq$ we see that the analytically continued $\det{\cal H}_k$ becomes just the (inverse) transmission coefficient for a wave of momentum -q hitting the potential $\varepsilon(z)$ from the right to propagate into the other the side of the potential. 
This is of course identical to the more conventionally defined (inverse) transmission coefficient for a wave of momentum +q coming from the left to propagate into the right of the potential. Denoting by t(q) the transmission coefficient and changing back from q to k we have the result: $\det{\cal H}_k=[t(ik)]^{-1}$.

Next we would like to study the case where the potential $\varepsilon(z)$ is generated by two objects one located around $z=0$ and the other around $z=a$ thus obtaining their mutual force.

Scattering in one dimension is usually described by a $2\times 2$ transfer matrix $T={\left(\begin{array}{cc} {t^*}^{-1}&-r^*{t^*}^{-1}\\ -rt^{-1}&t^{-1}\end{array}\right )}$ (satisfying $tt^*+rr^*=1$)\footnote{The ordinary S-matrix is of the form  $S ={\left(\begin{array}{cc} t&-r^*\\ r&t^*\end{array}\right )}$  }. A wave of the form $Ae^{iqz}+Be^{-iqz}$ coming from the left will be transformed upon crossing the scattering potential to the right into  $Ce^{iqz}+De^{-iqz}$ where ${\left(\begin{array}{c} C\\ D\end{array}\right )}=T{\left(\begin{array}{c} A\\ B\end{array}\right )}$.
As the location of the scattering potential is moved by a distance $a$ to the right without changing its shape the transfer matrix $T$ change into $A^{-1}TA$ where $A$ is the matrix representing translation: $A={\left(\begin{array}{cc} e^{iqa}&0\\ 0&e^{-iqa}\end{array}\right )}$.

For a potential $\varepsilon(z)$ generated by two different slabs one located around $z=0$ and the other around $z=a$ the total transfer matrix of the system is given by the product of the two transfer matrices corresponding to the two objects: $T_1$ and $A^{-1}T_2A$. so that we have $T=A^{-1}T_2AT_1$. And in particular $t^{-1}\equiv(T)_{2,2}=(T_1)_{1,2}(T_2)_{2,1}e^{2iqa}+(T_1)_{2,2}(T_2)_{2,2}={r_1^*r_2\over t_1^*t_2}e^{2iqa}+{1\over t_1t_2}$. Assuming real potential we have the relations $r(q)^*=r(-q),t(q)^*=t(-q)$ Which allow us to make the analytic continuation into $q=ik$ obtaining $\det{\cal H}_k={r_1(-ik)r_2(ik)\over t_1(-ik)t_2(ik)}e^{-2ka}+{1\over t_1(ik)t_2(ik)}$. 
For the Casimir force of interest acting between the bodies we need only the dependence on the distance $a$. Thus we may renormalize $\det{\cal H}_k$ into $1+{r_1(-ik)r_2(ik)t_1(ik)\over t_1(-ik)}e^{-2ka}$. Hence the Casimir interaction between the two bodies will be given by \beq E={1\over 2}\int{d^3k\over(2\pi)^3}\log(1+{r_1(-ik)r_2(ik)t_1(ik)\over t_1(-ik)}e^{-2ka})\eeq
For a massive field one obtains similar relation with $e^{-2ak}\rightarrow e^{-2a\sqrt{k^2+m^2}}$.\newline
We remark that upon replacing a potential $\varepsilon(z)$ by its mirror image  $\varepsilon(-z)$ the reflection coefficient $r$ changes into $\tilde{r}=-rt^*t^{-1}$ hence the last result may be also written as  \beq\label{gpp} E={1\over 2}\int{d^3k\over(2\pi)^3}\log(1-r_1(-ik)\tilde{r}_2(ik)e^{-2ka})\eeq which is explicitly symmetric under $T_1\leftrightarrow T_2^{-1}$ as expected.\footnote{If the two slabs are mirror image of eachother i.e $r_1\equiv\tilde{r}_2$ then eq(\ref{gpp}) imply that they would always attract}

As a simple example suppose that the scattering potential are given by delta-functions: $\varepsilon=\lambda_1\delta(z)+\lambda_2\delta(z-a)$. The transfer and reflection coefficient from a delta potential are $t(q)={2iq\over 2iq-\lambda},r(q)={\lambda\over 2iq-\lambda}$. Substituting this in the general result(\ref{gpp}) we have $ E={1\over 2}\int{d^3k\over(2\pi)^3}\log(1-{\lambda_1\over\lambda_1+2k}{\lambda_2\over\lambda_2+2k}e^{-2ka})$. 
In particular we see that for $\lambda\rightarrow\infty$ this reduces as expected to the standard Casimir energy of two parallel plates ${1\over 2}\int{d^3k\over(2\pi)^3}\log(1-e^{-2ka})=-{\pi^4\over 1440a^3}$. We can also see that when $\lambda$ becomes negative the energy develops an imaginary part as expected. It is also quite easy to generalize the last result to a situation where there are more then two plates Thus obtaining an explicit expression for a three body force which should exist as long as there is no perfect reflection from the middle plate. Also the case of several finite width slabs including circular periodic boundary conditions can be addressed by these methods\cite{bns}

The fact that the Casimir force may be expressed in terms of reflection and transmission coefficients seems very natural. It is known that one way for obtaining the Casimir force is by summing the pressure exerted on the various bodies as a result of the vacuum zero modes being scattered of them\cite{4}. Clearly such an approach should also result in expressing the force in terms of reflection and transmission coefficients.  
 
If instead of a single scalar we have a multicomponent field most of the above arguments can be repeated. The only difference being that now $t(k),r(k)$ are matrices in internal space and that we have to take the (finite dimensional) determinant also with respect to them. Thus $E/A=-{1\over 2}\int{d^3k\over(2\pi)^3}\log\det t(ik)$ in particular if the different components of the field do not mix the determinant becomes just the product of the various factors and the energy the sum of the different energies as expected. 
\footnote{Also for this case we can consider arrays of parallel slabs. For the case mentioned above of plates conducting in different directions which is relevant only for vectorial fields there is partial reflection and incomplete screening of nonadjacent plates}

In order to apply this to the Casimir force between two parallel dielectrics of dielectric constants $\varepsilon_1,\varepsilon_2$ and widths $D_1,D_2$ we just need to solve the simple classical problem of computing the transmission of a plane electromagnetic wave through the system made of these two plates. 
The classical transfer matrix for an electromagnetic wave when passing from one dielectric to another through the $z=0$ plane can be expressed quite simply in terms of the ratio $\epsilon_{1,2}\equiv\epsilon_1/\epsilon_2$ and the ratio $\alpha_{1,2}\equiv q_z^{(2)}/q_z^{(1)}=\sqrt{\epsilon_2\omega^2-q_x^2-q_y^2\over\epsilon_1\omega^2-q_x^2-q_y^2}$. For a wave whose polarization is parallel to the $z=0$ plane we have $T^{(\ominus)}={1\over 2\alpha}{\left(\begin{array}{cc} \alpha+1&\alpha-1 \\ \alpha-1&\alpha+1\end{array}\right )}$ while for the other polarization  $T^{(\oslash)}={1\over 2\alpha\epsilon}{\left(\begin{array}{cc} \alpha+\epsilon&\alpha-\epsilon \\ \alpha-\epsilon&\alpha+\epsilon\end{array}\right )}$. Let us also denote as before  $A(x)={\left(\begin{array}{cc} e^{ix}&0 \\ 0&e^{-ix}\end{array}\right )}$. If we have a system of two parallel dielectrics of widths $D_1,D_2$ and dielectric constants $\epsilon_1,\epsilon_2$ separated by a distance $a$ inside a medium of dielectric constant $\epsilon_0$ then the total transfer matrix through the system is given by \beq\label{trm}{T=A(-q_z(D_1+D_2+a))T(\epsilon_0,\epsilon_2)A(q_z^{(2)}D_2)T(\epsilon_2,\epsilon_0)A(q_za)T(\epsilon_0,\epsilon_1)A(q_z^{(1)}D_1)T(\epsilon_1,\epsilon_0)}\eeq The contribution of each polarization to the Casimir force between the two plates is $ {1\over 2}\int{d^3k\over(2\pi)^3}\log(T_{2,2}(q_z=ik))$. Since the two polarization are completly independent the total force is just the sum of these two contributions. It is more convinient to make the changes $\omega\rightarrow ik_t$ and $q_z\rightarrow ik$ already in the expressions (\ref{trm})for the matrix $T$ This amounts to the changes  $A(x)\rightarrow{\left(\begin{array}{cc} e^{-x}&0 \\ 0&e^x\end{array}\right )},\alpha\rightarrow\sqrt{\epsilon_2k_t^2+k_x^2+k_y^2\over\epsilon_1k_t^2+k_x^2+k_y^2}$ and possibly also $\epsilon(\omega)\rightarrow\epsilon(ik_t)$.

Substituting all this into eq(\ref{trm}) one easily finds that the corresponding functional determinant for the two polarizations (up to unimportant $a$-independent factors) are:
\beq\label{cfd} \det{\cal H}_k^{(\oslash)}=1-({\epsilon_1^2-\alpha_1^2\over\epsilon_1^2+\alpha_1^2+2\epsilon_1\alpha_1{\it coth}(\alpha_1 D_1)})({\epsilon_2^2-\alpha_2^2\over\epsilon_2^2+\alpha_2^2+2\epsilon_2\alpha_2{\it coth}(\alpha_2 D_2)})e^{-2ka}\eeq
$$ \det{\cal H}_k^{(\ominus)}=1-({1-\alpha_1^2\over 1+\alpha_1^2+2\alpha_1{\it coth}(\alpha_1 D_1)})({1-\alpha_2^2\over 1+\alpha_2^2+2\alpha_2{\it coth}(\alpha_2 D_2)})e^{-2ka}$$
(here we assumed $\epsilon_0=1$ and hence $\alpha_i=\sqrt{\epsilon_ik_t^2+k_x^2+k_y^2\over k_t^2+k_x^2+k_y^2}$)\newline The plate interaction energy is given (up to $a$-independent terms) by $E={1\over 2}\int{d^3k\over(2\pi)^3}(\log\det{\cal H}_k^{(\oslash)}+\log\det{\cal H}_k^{(\ominus)})$.

In particular if the widths $D_1,D_2$ are large (compared to $a$) then (\ref{cfd}) reduces to  $\det{\cal H}_k^{(\oslash)}=1-({\epsilon-\alpha\over\epsilon+\alpha})_{0,1}({\epsilon-\alpha\over\epsilon+\alpha})_{0,2}e^{-2ka}$ and $det{\cal H}_k^{(\ominus)}=1-({1-\alpha\over 1+\alpha})_{0,1}({1-\alpha\over 1+\alpha})_{0,2}e^{-2ka}$ for the two polarizations.  Substituting this into the general formula leads to the well known Lifshitz force\cite{lif}.

Equation (\ref{trm}) may also be used in more general scenarios e.g. to find exactly the torqe between two nonisotropic dielectric plates (evaluated in some approximation in\cite{enk}). The resulting expression for the energy may turn to be quite complicated in such a general case. Some simplification always occur for large  widths $D_1,D_2$ since than $A(q_zD)$ reduces in(\ref{cfd}) to the projection ${\left(\begin{array}{cc} 0&0 \\ 0&1\end{array}\right )}$
therefore ignoring $a$-independent factors we have  $T\propto\left(T(\epsilon_2,\epsilon_0)
{\left(\begin{array}{lc} e^{-2ka}&0 \\ 0&1\end{array}\right )}T(\epsilon_0,\epsilon_1)\right)_{22}$

The above results reduce the problem of computing the Casimir energy for general plate geometry to the problem of computing the reflection coefficient from the corresponding one dimensional potential. Generalizing this method to arbitrary non planar geometry would be very difficult if not impossible. However one may still try to use similar method for problems of spherical symmetry. Using partial wave expansion reduces this to an essentially one dimensional problem. Thus we may identify (each specific partial wave contribution to) the functional determinant with the ratio of the outgoing to the ingoing amplitude i.e. with the phase shift or more accurately with $e^{2i\delta_l}$.
This consideration leads us immediately to the following exact formal expression for the Casimir energy\beq\label{edl} E=\sum_l(2l+1)\int_0^\infty{dk\over 2\pi}\delta_l(k)\eeq This result also follows imeadiatly from known relation between phase shifts and density of states $n(E)={1\over\pi}{d\delta\over dE}$.
The same result can also be obtained by more standard considerations. Suppose we want to compute the Casimir energy by explicitly finding the eigenfrequncies and summing ${1\over 2}\sum\hbar\omega$. 
Since the asymptotic form of the partial wave at infinity is $\sim{1\over r}\sin(kr-\delta_l)$ we easily see that enforcing a boundary condition stating that the fields should vanish on some sphere of very large radius $R$ leads to the simple equation $kR-\delta_l(k)=n\pi$ for the eigenvalues.
The Casimir energy of a specific partial wave is therefore $E_l={1\over 2}\sum_n k={1\over 2}\sum{n\pi+\delta_l(k)\over R}=const+{1\over 2R}\sum\delta_l(k)$. In this equation $k=k(n)$ is the solution of $kR-\delta_l(k)=n\pi$ however since
the diameter $a$ of the scattering potential is much smaller then $R$ we always have $kR>>\delta_l(k)$ and therefore we may write simply $k={n\pi\over R}$. As $R>>a$ the sum over n may be replaced by an integral $\sum_n\rightarrow R\int_0^\infty{dk\over \pi}$ leading us back to eq(\ref{edl}). However equation(\ref{edl}) unlike(\ref{gpp}) is in general divergent and needs a nontrivial regularization. The difficulty originates from the fact that our previous normalization condition that of demanding the energy to vanish as the different bodies are moved infinitly apart from eachother without changing their shape becomes meaningless here.

An issue related to the above which we elaborate elsewhere\cite{mn1} is the question of experimental testing the repulsive nature of the Casimir energy of the sphere\cite{boy,sch}. One cannot contrary to what has been (implied/)stated by various authours\cite{em}\cite{lam,mjh} check this by finding that two near by conducting hemispheres repel. Indeed the latter case of saperated hemispheres represents a well defined and in principle testable experimental set-up. However such experiment are likely to manifest attraction of the two hemispheres. Thus consider the two thin rims at the boundaries of the hemispheres which are a small distance $\delta$ apart. On their own the rims have an attractive Casimir attraction of $F_{rims}\sim 2\pi R{\hbar c\over\delta^3}$. Since the wavelength of vacuum fluctuations which generate most of this attraction are $\lambda\approx\delta$, the existance of the remaining bulk of the hemispheres of radius $R\gg\delta$ is not expected to drastically modify this force. Since this force is vastly larger (by $(R/\delta)^3$!) than the repulsive force of a single sphere $F_{sphere}\sim{\hbar c\over R^2}$ the attractive nature is not modified.

\vspace{4mm}{\bf Acknowledgment} I thank Prof S.Nussinov for discussions and his  interest.


\newpage
\begin{thebibliography}{99}


\bibitem{cas1} H.B.G. Casimir,{\em ''on the attraction between two perfectly conducting plates''} Proc. K.Ned. Akad.Wet. 51, 793-796(1948)

\bibitem{rr} V.M.Mostepanenko, N.N.Trunov, {\em The Casimir effect and its application} Clarendon press oxford 1997

\bibitem{11}O.Kenneth and S.Nussinov, preprint hep-th/9802149.

\bibitem{lif}E.M.Lifshitz, Sov.Phys.JETP 2,73(1956)

\bibitem{col}S.Coleman {\em ``Aspects of symmetry''} Cambridge University Press (page 340)

\bibitem{boy} T.H.Boyer ,Phys.Rev.174,1764(1968)

\bibitem{sch}J.Schwinger,L.L.DeRaad,K.A.Milton,Ann. Phys.(N.Y)115,338(1978)

\bibitem{cyl}L.L.DeRaad.and K.A.Milton,Ann. Phys.(N.Y)136,229(1981)

\bibitem{lam} S.K.Lamoreaux, {\em ``Demonstration of the Casimir force in the 0.6 to 6$\mu m$ Range''},Phys Rev. Lett. 78(1) 5-8(1997)

\bibitem{mjh}U.Mohideen, and Anushree Roy. preprint physics/9805038
\newline Anushree Roy,Chiung-Yuan Lin and U.Mohideen quant-ph/9906062 Phys. Rev. D60:111101(1999)

\bibitem{enk}S.J.Van Enk, Phys Rev A52(4) 2569-2575(1995)

\bibitem{4} P.W.Miloni, R.J.Cook and M.F.Goggin, Phys.Rev A38, 1621(1988)
 
\bibitem{em}E.Elizalde and A.Romeo, Am.J.Phys. Vol.59.No.8 711(1991)

\bibitem{bns} I.Brevik, H.B.Nielsen, S.D.Odintsov Phys.Rev D53 3224(1996)
\newline I.Brevik, R.Sollie J.Math.Phys.38(6) 2774(1997)
\newline M.H.Berntsen, I.Brevik Ann Phys 257(1) 84(1997)


\bibitem{mn1}O. Kenneth,S.Nussinov{\em Small object limit of Casimir effect and signs of the Casimir force}(to appear next week)

\bibitem{n2}O.Kenneth, S.Nussinov {\em On possibility experimental verification of the newly suggested polarized version of the Casimir effect} (to appear soon)

\end{thebibliography}
\end{document}